\documentclass[10pt]{article}
\usepackage[T1]{fontenc}
\usepackage[latin9]{inputenc}
\pdfoutput=1
\usepackage{xcolor}
\usepackage{pdfcolmk}
\usepackage{amsmath}
\usepackage{amssymb}
\PassOptionsToPackage{normalem}{ulem}
\usepackage{ulem}

\makeatletter

\providecolor{lyxadded}{rgb}{0,0,1}
\providecolor{lyxdeleted}{rgb}{1,0,0}

\usepackage{enumitem}		

\usepackage{M11}

\renewcommand{\bar}[1]{\oldbar{#1}}

\usepackage{slashed}
\renewcommand{\not}[1]{\slashed{#1}}

\AtBeginDocument{
  
}

\makeatother

\begin{document}
\setpreprint{arXiv:\href{http://arxiv.org/abs/1509.07872}{1509.07872} \\ {\footnotesize QGASLAB}-15-06}

\title{Fermionic T-Duality of $AdS_{n}\times S^{n}\:(\times S^{n})\times T^{m}$
\\
using IIA Supergravity }

\author{Michael C. Abbott, Jeff Murugan, \& Justine Tarrant~ \address{ 
The Laboratory for Quantum Gravity \& Strings,\\
Department of Mathematics \& Applied Mathematics, \\ University of Cape Town, Rondebosch 7701, South Africa}}

\date{
September 2015}
\maketitle
\begin{abstract}
We show that the string backgrounds $AdS_{2}\times S^{2}\times T^{6}$
and $AdS_{d}\times S^{d}\times S^{d}\times T^{10-3d}$ ($d=2,3$)
are self-dual under a series of bosonic and fermionic T-dualities.
We do this using the fermionic Buscher rules derived by Berkovits
and Maldacena, thus working at the level of the supergravity fields.
This allows us to explicitly track the behaviour of the RR fields,
from which we see that we need T-duality along some torus directions.
For the $AdS\times S\times S$ cases, which contain cosets of $D(2,1;\alpha)^{d-1}$,
it is necessary to perform bosonic T-duality along some complexified
Killing spinors in one of the spheres. 
\end{abstract}
\newcommand{\cE}{\mathcal{E}}
\newcommand{\ket}[1]{\left\vert #1 \right\rangle}
\newcommand{\C}{\mathrm{C}}

\section{Introduction}

An important theme in our understanding of theAdS/CFT correspondence
\cite{Maldacena:1997re} has been that both theories have far more
symmetry than initially meets the eye. The largest example of this
was the discovery of integrability on both sides of the correspondence,
giving an infinite tower of conservation laws \cite{Beisert:2010jr}.
But a particularly important example is that of dual conformal symmetry,
which began life as a symmetry of scattering amplitudes of $\mathcal{N}=4$
SYM \cite{Drummond:2008vq}, acting on the momenta in the same way
that ordinary conformal symmetry acts on positions. The AdS/CFT connection
between scattering amplitudes and Wilson loops is built on this symmetry,
associating each amplitude to a string worldsheet in a dual AdS space.
The symmetry extends to the whole superconformal group, and it is
believed to be part of the structure of integrability \cite{Beisert:2009cs,Drummond:2010km}. 

The existence of dual conformal symmetry was later understood to be
a consequence of the self-duality of the string background $AdS_{5}\times S^{5}$
\cite{Berkovits:2008ic,Beisert:2008iq}, meaning that the total effect
of a sequence of bosonic and fermionic T-duality transformations is
to return us to exactly the same background. The bosonic dualities
along boundary directions of $AdS_{5}$ map the momenta of the scattering
amplitude into the path bounding the Wilson loop. Fermionic T-duality
is a cousin of Bosonic T-duality, acting along a Killing spinor, or
in fact necessarily a complexified Killing spinor, rather than a Killing
vector. It was introduced by \cite{Berkovits:2008ic,Beisert:2008iq};
for reviews see \cite{OColgain:2012si,Bakhmatov:2011ab} and for other
perspectives see \cite{Beisert:2009cs,Sfetsos:2010xa}. As in the
bosonic case, one adds a constrained auxiliary field and then integrates
out the original field (i.e. original target space co-ordinate, or
fermion) to obtain an action of the same shape which can be interpreted
as describing strings on a dual background. Rather than performing
this process for each background we encounter, we can do it once and
write down the rules for mapping one background to another, usually
called the Buscher rules \cite{Buscher:1987qj}. The rules for bosonic
T-duality are now understood to be particular generators of the $O(D,D)$
symmetry of generalised geometry \cite{Alvarez:1994dn}. 

The correspondence between $\mathcal{N}=4$ SYM and IIB strings in
$AdS_{5}\times S^{5}$ is by far the best understood example of AdS/CFT.
We are interested in extending this understanding to other examples,
with less than maximal supersymmetry. In the correspondence between
ABJM theory and IIA strings on $AdS_{4}\times CP^{3}$ \cite{Aharony:2008ug,Klose:2010ki},
dual conformal symmetry is observed in the amplitudes \cite{Huang:2010qy,Lipstein:2011zz,Bianchi:2012cq}
but attempts to show T-self-duality in the string theory have failed
\cite{Adam:2009kt,Adam:2010hh,Bargheer:2010hn,Bakhmatov:2010fp},
apart from the much simpler case of the pp-wave limit \cite{Bakhmatov:2011aa}.
We have nothing to add to this puzzle now, but turn instead to $AdS_{3}$
and $AdS_{2}$ backgrounds. 

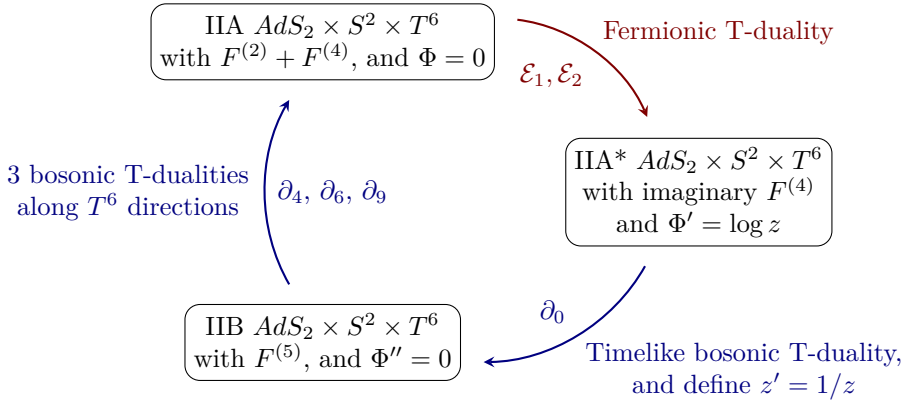
\begin{figure}
\centering 

\begin{tikzpicture}[scale=1,rounded corners=2mm, auto,bend left, shorten >= 3mm,shorten <= 3mm, >=stealth] 


\node (top) [rectangle,draw,align=center] at (0,4) {IIA $AdS_2\times S^2\times T^6$ \\ with $F^{(2)} + F^{(4)}$, and $\Phi=0$};

\node (right) [rectangle,draw,align=center] at (5,2) {IIA* $AdS_2\times S^2\times T^6$ \\ with imaginary $F^{(4)}$ \\ and $\Phi'=\log z$};

\node (bot) [rectangle,draw,align=center] at (0,0) {IIB $AdS_2\times S^2\times T^6$ \\ with $F^{(5)}$, and $\Phi''=0$};

\draw [->,thick,darkred] (top) to node {Fermionic T-duality} node [swap] {$\cE_1,\cE_2$} (right);

\draw [->,thick,darkblue] (right) to node [align=center] {Timelike bosonic T-duality, \\ and define $z'=1/z$} node [swap] {$\partial_0$} (bot);

\draw [->,thick,darkblue] (bot) to node [align=center] {3 bosonic T-dualities \ \\ along $T^6$ directions \ } node [swap] {$\partial_4$, $\partial_6$, $\partial_9$} (top);


\end{tikzpicture}

\caption{The idea of self-duality we study is that a sequence of fermionic
and bosonic T-dualities returns us to the same background. This is
depicted here for the case in which we start with IIA $\protect\smash{AdS_{2}\times S^{2}\times T^{6}}$
supported by $F^{(2)}$ and $F^{(4)}$ RR fields (section \ref{sub:Aside-on-F2-F4-case}).
\label{fig:IIA-diagram}}
\end{figure}

The study of strings in $AdS_{5}\times S^{5}$ was greatly facilitated
by the observation that the Green--Schwarz (GS) action can be written
using a $\mathbb{Z}_{4}$-graded coset structure (i.e. the supersymmetric
cousin of a symmetric space) $PSU(2,2\vert4)/SO(1,4)\times SO(5)$
\cite{Metsaev:1998it,Cagnazzo:2012uq}. This idea was used heavily
in integrability \cite{Bena:2003wd,Beisert:2005bm,SchaferNameki:2010jy},
and also in the understanding of fermionic T-duality: Beisert, Ricci,
Tseytlin and Wolf \cite{Beisert:2008iq} work entirely with the coset
action, fixing a particular $\kappa$-symmetry gauge at the start,
and find that the dual model amounts to a different choice of $\kappa$-gauge.
Similar cosets are useful in describing strings on $AdS_{4}$, $AdS_{3}$
and $AdS_{2}$ backgrounds, but there are more caveats. The coset
$OSP(6|4)/SO(2,1)\times U(3)$ describes a partially $\kappa$-fixed
GS string in $AdS_{4}\times CP^{3}$, although this gauge choice forbids
some solutions of interest \cite{McLoughlin:2008he}. 

Fermionic T-duality of $AdS_{3}\times S^{3}\times T^{4}$ (with RR
flux) was studied using the coset model $PSU(1,1\vert2)^{2}/SU(1,1)\times SU(2)$
by Adam, Dekel and Oz \cite{Adam:2009kt}. This describes GS strings
in a $\kappa$-symmetry gauge which turns off all the non-coset fermions;
like the gauge used to write $AdS_{4}\times CP^{3}$ as a coset, this
one is not always legal. Another approach was explored by \'{O} Colg\'{a}in
\cite{OColgain:2012ca}, using the Buscher rules derived by Berkovits
and Maldacena \cite{Berkovits:2008ic}. This avoids all mention of
both cosets and $\kappa$-symmetry, working only with the supergravity
fields of the background. It also allows one to see the effect on
the RR fields explicitly. (This was used for $AdS_{5}\times S^{5}$
in \cite{Berkovits:2008ic}; see also \cite{Bakhmatov:2009be,Godazgar:2010ph,Bakhmatov:2010fp}
for other applications.) 

In this paper we use this idea of working only with the supergravity
fields to prove the self-duality of backgrounds $AdS_{2}\times S^{2}\times T^{6}$
and $AdS_{d}\times S_{+}^{d}\times S_{-}^{d}\times T^{10-3d}$ for
$d=2,3$. We find it more convenient to begin in the type IIA theory,
although the bosonic T-dualities necessarily take us through IIB.
(See figure \ref{fig:IIA-diagram}.) For the cases with $S_{+}^{d}\times S_{-}^{d}$
we show that it is necessary to perform bosonic T-duality along some
complexified Killing vectors of one of the spheres. 

There is some overlap between this paper and another longer paper
\cite{Abbott:2015mla}, written as part of a larger collaboration.
The presentation of the supergravity approach there is however very
different, using the currents of the coset model to generate Killing
spinors, thus treating many backgrounds at once. The aim of the present
paper is to give a much more elementary (if less general) explanation
of the duality of these backgrounds.

\subsection*{Outline}

In section \ref{sec:Preliminaries} we fix some notation and recall
the Buscher rules for fermionic T-duality. Section \ref{sec:AdS2-Backgrounds}
treats backgrounds $AdS_{2}\times S^{2}\times T^{6}$ (for two different
choices of RR flux) and $AdS_{2}\times S^{2}\times S^{2}\times T^{4}$
(which requires bosonic T-duality in a complex direction). Section
\ref{sec:AdS3-Backgrounds} repeats this one dimension up: $AdS_{3}\times S^{3}\times S^{3}\times S^{1}$.
We conclude in section \ref{sec:Conclusions}. 

Appendix \ref{sec:Bosonic-T-duality} reviews the necessary facts
about bosonic T-duality (including a nontrivial dilaton). Appendix
\ref{sec:Gamma-Matrices} has gamma matrix conventions.

\section{Preliminaries\label{sec:Preliminaries}}

In type IIA string theory we have two Majorana--Weyl spinors of opposite
chirality. Since we use a basis in which $\Gamma_{11}=\sigma^{3}\otimes1$,
we can write these as 
\[
\begin{pmatrix}\epsilon^{\beta}\\
0
\end{pmatrix}\quad\mbox{and}\quad\begin{pmatrix}0\\
\hat{\epsilon}_{\beta}
\end{pmatrix},\qquad\beta=1\ldots16.
\]
It is convenient to write the Killing spinors using one 32-dimensional
$E=(\epsilon,\hat{\epsilon})$. Our notation is that $\mu,\nu=t,x,z,\theta_{+}\ldots\psi_{-},u$
are curved and $m,n=0,1,2\ldots9$ flat spacetime indices, and $\alpha,\beta$
spinor indices as above. The constraints on a Killing spinor $E$
from the gravitino and dilatino variations are 
\begin{equation}
D_{\mu}E=-\frac{1}{8}S\Gamma_{\mu}E,\qquad TE=0\label{eq:killing-eq-wulff}
\end{equation}
where the covariant derivative is $D_{\mu}E=\partial_{\mu}E+\frac{1}{4}\omega_{\mu np}\Gamma^{np}E$,
and we define

\[
S\equiv\not F^{(2)}\Gamma_{11}+\not F^{(4)},\qquad T\equiv\frac{i}{16}\Gamma_{m}S\Gamma^{m}=\frac{3i}{8}\not F^{(2)}\Gamma_{11}+\frac{i}{8}\not F^{(4)}.
\]

Fermionic T-duality acts on the dilaton and the RR forms, but leaves
the metric and the NS-NS two-form invariant. We will employ here the
Buscher rules for this derived by Berkovits and Maldacena \cite{Berkovits:2008ic}.
The direction along which we dualise is specified by a complex Killing
spinor, which we write $\cE=E_{1}+iE_{2}$ etc. We can perform the
duality along several directions $\cE_{i}$ at once. These must obey
an orthogonality condition 
\begin{equation}
\bar{\cE}_{i}\Gamma_{\mu}\cE_{j}=0\:\forall\mu,i,j,\qquad\mbox{where }\bar{\cE}\equiv\cE^{T}\Gamma^{0}\label{eq:rules-orthog}
\end{equation}
which is not solved by any real spinors---this is why we must use
complex Killing spinors. Then we find a matrix $\C$ from 
\begin{equation}
\partial_{\mu}\C_{ij}=i\bar{\cE}_{i}\Gamma_{\mu}\Gamma_{11}\cE_{j}\label{eq:rules-dC}
\end{equation}
 and this gives the change in the dilaton\footnote{Notice that scaling any $\cE_{i}$ by a constant factor changes $\Phi'$
by a constant, and has no effect on $\Delta F$. } 
\begin{equation}
\Delta\Phi\equiv\Phi'-\Phi=\frac{1}{2}\log(\det\C)\label{eq:rules-deltaPhi}
\end{equation}
and in the RR forms 
\begin{equation}
\Delta F\equiv e^{\Phi'}F_{\,\;\;\beta}^{\prime\alpha}-e^{\Phi}F_{\;\;\beta}^{\alpha}=\frac{16}{i}\,(\C^{-1})_{ij}\:\epsilon_{i}^{\alpha}\:\hat{\epsilon}_{j\beta}.\label{eq:rules-deltaF}
\end{equation}
Here the lower-case spinors are now complex, $\cE_{i}=(\epsilon_{i},\hat{\epsilon}_{j})$,
and $F$ is the bispinor encoding the RR forms, defined 
\begin{equation}
F_{\;\;\beta}^{\alpha}\equiv\frac{1}{2!}\:F_{mn}^{(2)}\:(\gamma^{m})^{\alpha\gamma}(\gamma^{n})_{\gamma\beta}+\frac{1}{4!}\:F_{mnpq}^{(4)}\:(\gamma^{m}\gamma^{n}\gamma^{p}\gamma^{q})_{\;\:\beta}^{\alpha}.\label{eq:IIA-bispinor-defn}
\end{equation}
The lower-case gamma matrices are the following $16\times16$ blocks
(full details are in appendix \ref{sec:Gamma-Matrices})
\[
\Gamma^{m}=\left[\begin{array}{cc}
0 & (\gamma^{m})^{\alpha\beta}\\
(\gamma^{m})_{\alpha\beta} & 0
\end{array}\right].
\]

\section{$AdS_{2}\times M$ Backgrounds\label{sec:AdS2-Backgrounds}}

This section looks at $AdS_{2}\times S^{2}\,(\,\times S^{2})\times T$
backgrounds, all having 8 supersymmetries. However we are interested
only in those supersymmetries which commute with the $\partial AdS$
translations along which we will perform bosonic T-duality. This restricts
us to considering the 4 Poincar\'e Killing spinors, allowing for
2 complex directions along which to perform fermionic T-duality. 

The metric has a parameter $\alpha\in[0,1]$, such that at $\alpha=1$
the second sphere becomes flat: 
\begin{equation}
ds^{2}=ds_{AdS}^{2}+\frac{1}{\alpha}ds_{S_{+}}^{2}+\frac{1}{1-\alpha}ds_{S_{-}}^{2}+\sum_{i}dy_{i}^{2}\quad=\eta_{mn}e_{\mu}^{m}e_{\nu}^{n}dx^{\mu}dx^{\nu}.\label{eq:defn-metric-1}
\end{equation}
We adopt Poincar\'e co-ordinates for $AdS_{2}$
\[
ds_{AdS_{2}}^{2}=\frac{-dt^{2}+dz^{2}}{z^{2}},\qquad\Rightarrow\quad\omega_{t[01]}=\frac{1}{z}
\]
and for the spheres,
\begin{equation}
ds_{S_{\pm}^{3}}^{2}=d\theta_{\pm}^{2}+\sin^{2}\theta_{\pm}d\varphi_{\pm}^{2},\qquad\Rightarrow\quad\omega_{\varphi[23]}=-\cos\theta_{+},\quad\omega_{\varphi_{-}[45]}=-\cos\theta_{-}\label{eq:metric-S2}
\end{equation}
and similarly for $S_{-}^{3}$. Note that the spin connection components
$\omega_{\mu ab}$ are independent of $\alpha$ (although the vielbeins
$e_{\mu}^{m}$ are of course not). We take $\Phi=0$, the following
RR flux: 
\begin{equation}
\not F^{(4)}=\Gamma^{01}\left(-\Gamma^{67}+\Gamma^{89}\right)+\sqrt{\alpha}\Gamma^{23}\left(\Gamma^{68}+\Gamma^{79}\right)+\sqrt{1-\alpha}\Gamma^{45}\left(-\Gamma^{78}+\Gamma^{69}\right).\label{eq:AdS2-F4}
\end{equation}
At $\alpha=1$, bosonic T-duality along the $x^{5}$ direction takes
us to the IIB $AdS_{2}\times S^{2}\times T^{6}$ case studied by \cite{Sorokin:2011rr},
with only $F^{(5)}$. (Section \ref{sub:Aside-on-F2-F4-case} below
looks at an alternative IIA choice.) Following \cite{Wulff:2014kja}
we can write 
\[
\not F^{(4)}=-4\Gamma^{0167}P_{1}\left(1-P_{2}\right)
\]
where 
\[
P_{1}=\frac{1}{2}\left(1+\Gamma^{6789}\right),\qquad P_{2}=\frac{1}{2}\left(1+\sqrt{\alpha}\Gamma^{012378}+\sqrt{1-\alpha}\Gamma^{014568}\right).
\]

The Killing spinor equation is $D_{\mu}E=-\frac{1}{8}\not F^{(4)}\Gamma_{\mu}E$,
and we can build up a solution $E=z^{-1/2}R_{S+}R_{S-}\xi$ as follows:
\begin{itemize}
\item We look for Poincar\'e killing spinors, i.e. those independent of
the boundary co-ordinates of $AdS$. Here this means $\partial_{t}E=0$,
and thus the rest of the $\mu=t$ equation gives us a constraint
\[
E=-\Gamma^{067}P_{1}P_{2}E.
\]
The $\mu=z$ equation then reads $\partial_{z}E=-1/2z\:E$, with solution
$E\propto z^{-1/2}$.
\item When $\mu\in S_{+}$ we get (using $P_{1}E=P_{2}E=E$) 
\begin{equation}
D_{\mu}E=\frac{\sqrt{\alpha}}{2}\Gamma_{\mu}\Gamma^{2368}E.\label{eq:AdS2-sphere-eq}
\end{equation}
L\"{u}, Pope and Rahmfeld \cite{Lu:1998nu} gave the Killing spinors
for $S^{2}$ as follows: the equation $D_{\mu}\epsilon=\pm\frac{i}{2}e_{\mu}^{a}\Gamma_{a}\epsilon$
is solved by $\epsilon=\exp(\pm i\frac{\theta}{2}\Gamma^{2})\exp(\frac{\varphi}{2}\Gamma^{23})$.
The $\theta$ equation is trivial, and the $\varphi$ equation uses
$(i\Gamma^{2})^{2}=-1$ and $\{i\Gamma^{2},\Gamma^{23}\}=0$. Our
equation replaces $i\Gamma^{2}\to\Gamma^{368}$ which still satisfies
these conditions. We also have radius $1/\sqrt{\alpha}$ which does
not affect the solution: 
\[
R_{S+}=\exp\Big(\frac{\theta_{+}}{2}\,\Gamma^{368}\Big)\exp\Big(\frac{\varphi_{+}}{2}\,\Gamma^{23}\Big).
\]

\item When $\mu\in S_{-}$ we get instead
\[
D_{\mu}E=\frac{\sqrt{1-\alpha}}{2}\Gamma_{\mu}\Gamma^{4578}E.
\]
Notice that $[\Gamma_{\mu}\Gamma^{4578},R_{S+}]=[D_{\mu},R_{S+}]=0$,
so this is precisely the equation which must be solved by $R_{S-}$
alone. 
\end{itemize}
The complete solution is then 
\begin{equation}
E(\xi)=\frac{1}{\sqrt{z}}\:\exp\Big(\frac{\theta_{+}}{2}\,\Gamma^{368}\Big)\exp\Big(\frac{\varphi_{+}}{2}\,\Gamma^{23}\Big)\:\exp\Big(\frac{\theta_{-}}{2}\,\Gamma^{578}\Big)\exp\Big(\frac{\varphi_{-}}{2}\,\Gamma^{45}\Big)\:\xi\label{eq:AdS2-killing}
\end{equation}
with constant $\xi$ obeying 
\begin{equation}
-\Gamma^{067}\xi=P_{1}\xi=P_{2}\xi=\xi\label{eq:AdS2-xi-constraint}
\end{equation}
(since the $R_{S\pm}$ commute with these conditions). There are exactly
4 such vectors $\xi_{a}$, and we may take them to be orthogonal and
normalised: $\xi_{a}\cdot\xi_{b}=\delta_{ab}.$

Fermionic T-duality needs complex combinations $\cE_{i}$ of these
Killing spinors. In section \ref{sub:AdS2-exceptional-case} we will
want to explore the most general possibility, so let us write these
as\newcommand{\myM}{\text{V}} \newcommand{\myN}{\text{W}} 
\[
\cE_{i}=b_{ia}E(\xi_{a}),\qquad a=1\ldots4.
\]
Then the orthogonality condition \eqref{eq:rules-orthog} reads 
\begin{equation}
b_{ia}\myM_{ab}^{\mu}b_{jb}=0,\qquad\myM_{ab}^{\mu}\equiv\bar{E}_{a}\Gamma^{\mu}E_{b}\,.\label{eq:defn-V-orthog}
\end{equation}
In the present $AdS_{2}$ case it is easy to show (using $\Gamma^{067}E=-E$)
that $\myM^{\mu}=0$ except for $\mu=t$, where 
\[
\myM_{ab}^{t}=E_{a}^{T}\Gamma^{0}z\Gamma^{0}E_{b}=-\xi_{a}^{T}\xi_{b}=-\delta_{ab}.
\]
Thus the orthogonality condition on the coefficients $b_{ia}$ is
simply $b_{ia}b_{ja}=0$.

\subsection{Case $\alpha=1$: $AdS_{2}\times S^{2}\times T^{6}$ \label{sub:AdS2-simplest-case}}

For the simplest case we can proceed almost exactly parallel to \cite{OColgain:2012ca}'s
treatment of IIB $AdS_{3}\times S^{3}\times T^{4}$, although it is
even simpler as we have only two complexified Killing spinors. 

Using $P_{2}E=E$ the $\mu\in S_{+}$ equation \eqref{eq:AdS2-sphere-eq}
above simplifies to $D_{\mu}E=\frac{1}{2}\Gamma_{\mu}\Gamma^{1}E$,
and thus the solution is just 
\begin{equation}
E(\xi)=\frac{1}{\sqrt{z}}\:\exp\Big(\frac{\theta}{2}\,\Gamma^{21}\Big)\exp\Big(\frac{\varphi}{2}\,\Gamma^{23}\Big)\:\xi\label{eq:AdS2-simple-killing}
\end{equation}
with $\xi_{a}$, $a=1\ldots4$ still solving \eqref{eq:AdS2-xi-constraint}:\footnote{i.e. each $\xi$ has four nonzero entries at the positions indicated,
with $\beta=1\ldots32$ by a temporary abuse of notation. } 
\[
(\xi_{a})^{\beta}=\frac{1}{2}\delta_{2a-\beta-1}+\frac{1}{2}\delta_{2a-\beta+8}+\frac{1}{2}\delta_{2a-\beta+16}-\frac{1}{2}\delta_{2a-\beta+23}.
\]
Choosing the complex combinations $\cE_{i}=b_{ia}E(\xi_{a})$ specified
by 
\[
b_{1}=(1,0,i,0),\qquad b_{2}=(0,1,0,i)
\]
and integrating \eqref{eq:rules-dC}, we get 
\[
\C=\frac{1}{z}\left[\begin{array}{cc}
i\cos\theta_{+}-\sin\theta_{+}\sin\phi_{+} & -\cos\phi_{+}\sin\theta_{+}\\
-\cos\phi_{+}\sin\theta_{+} & i\cos\theta_{+}+\sin\theta_{+}\sin\phi_{+}
\end{array}\right].
\]
Taking the determinant, \eqref{eq:rules-deltaPhi} gives the shift
in the dilaton to be 
\begin{equation}
\Delta\Phi=\Phi'-\Phi=-\log(z).\label{eq:AdS2-dilaton-fermishift}
\end{equation}
Bosonic T-duality along time (the only boundary direction of $AdS_{2}$)
un-does this shift, giving $\Phi''-\Phi'=\log z$. It also changes
the metric as follows: 
\begin{equation}
ds_{AdS}^{2}=\frac{-dt^{2}+dz^{2}}{z^{2}}\;\to\;-z^{2}\,dt^{2}+\frac{dz^{2}}{z^{2}}=\frac{-dt^{2}+dz^{\prime2}}{z^{\prime2}}\label{eq:AdS2-metric-zprime}
\end{equation}
where in the last step we define $z'=1/z$ to bring the metric back
to its original form. 

Turning to the RR forms, $\Delta F$ from \eqref{eq:rules-deltaF}
has real terms cancelling the original flux \eqref{eq:AdS2-F4}, and
imaginary terms leading to 
\[
\begin{aligned}e^{\Phi'}\not F^{(2)\prime} & =i\Gamma^{14}\\
e^{\Phi'}\not F^{(4)\prime} & =-i\Gamma^{0235}-i\Gamma^{1569}+i\Gamma^{1578}.
\end{aligned}
\]
Then bosonic duality along directions $x^{0}$, $x^{4}$, $x^{6}$
and $x^{7}$ returns us to the original RR flux. Notice that we need
to dualise along half of the torus directions here, and that the total
number of bosonic dualities is four (as it was in $AdS_{5}\times S^{5}$).\footnote{This was also seen in \cite{OColgain:2012ca}'s supergravity treatment
$AdS_{3}\times S^{3}\times T^{4}$, where it was necessary to dualise
along two torus directions and the two boundary directions of $AdS_{3}$. }

The earlier treatment of $AdS_{2}\times S^{2}$ in \cite{Adam:2009kt,Dekel:2011qw}
studied only the coset $\sigma$-model $PSU(1,1\vert2)/U(1)^{2}$,
rather than the whole critical string theory. Thus the need for bosonic
T-duality along torus directions was not seen there. This approach
was extended in \cite{Abbott:2015mla} to include non-coset fermions,
which forced us to include duality along torus directions.

\subsection{Case $\alpha<1$: $AdS_{2}\times S^{2}\times S^{2}\times T^{4}$\label{sub:AdS2-exceptional-case}}

Using the setup for generic complex combinations $\cE_{i}=b_{ia}E_{a}$
above, it is fairly easy to prove that no choice of $b_{i,a}$ can
lead to $\Delta\Phi=\log z+\mbox{const.}$ Fermionic T-duality thus
necessarily produces a shift in the dilaton which depends on some
of the sphere co-ordinates $\theta_{\pm},\varphi_{\pm}$, which bosonic
T-duality along time (and flat torus directions) cannot undo. 

\newcommand{\mylam}{\lambda_3}\newcommand{\myisom}{\lambda_{+}}

The way around this problem is to allow bosonic T-duality along a
complex Killing vector. To do this we introduce some unusual co-ordinates
for $S^{2}$ which arise from the form of the coset element used in
\cite{Abbott:2015mla}.\footnote{We suppress all except this $SU(2)$ part. We have also re-scaled
the algebra to avoid factors $c=\sqrt{\alpha}$. See also \cite{Dekel:2011qw}
for discussion of some related parameterisations. } There, $S^{2}=SU(2)/U(1)$ is parameterised by 
\begin{equation}
g=e^{\lambda_{+}L_{+}}e^{-\lambda_{3}L_{3}}\in SU(2),\qquad L_{\pm}=iL_{1}\pm L_{2},\qquad L_{n}=\frac{1}{2i}\sigma_{n}\label{eq:SS-group-element}
\end{equation}
with the coset's denominator taken to correspond to the first factor
in $\mathfrak{su}(2)=\mathfrak{g}_{(0)}\oplus\mathfrak{g}_{(2)}$,
where 
\[
\mathfrak{g}_{(0)}=\left\langle L_{+}+L_{-}\right\rangle =\left\langle L_{1}\right\rangle ,\qquad\mathfrak{g}_{(2)}=\left\langle L_{+}-L_{-},L_{3}\right\rangle =\left\langle L_{2},L_{3}\right\rangle .
\]
Then writing $J_{(2)}$ for the projection of the current onto $\mathfrak{g}_{(2)}$:
\[
J_{(2)}=\left[g^{-1}dg\right]_{(2)}=e^{i\lambda_{3}}d\lambda_{+}L_{2}+d\lambda_{3}L_{3}
\]
the metric is 
\begin{equation}
ds_{S_{+}}^{2}=-2\Tr\left(J_{(2)}J_{(2)}\right)=d\mylam^{2}+e^{2i\mylam}d\myisom^{2}.\label{eq:metric-S2-lambda}
\end{equation}
Comparing this to \eqref{eq:metric-S2}, it is easy to show that the
new co-ordinates $x^{\underline{2}}=\mylam$ and $x^{\underline{3}}=\myisom$
are related to the original ones by 
\begin{equation}
\begin{aligned}e^{i\mylam} & \equiv\cos\theta_{+}+i\sin\theta_{+}\cos\varphi_{+}\\
\myisom & \equiv e^{-i\mylam}\sin\theta_{+}\sin\varphi_{+}=\frac{\sin\varphi_{+}}{\cot\theta_{+}+i\cos\varphi_{+}}.
\end{aligned}
\label{eq:SS-complex-coord}
\end{equation}
Below we will write $e^{3}=e_{\mu}^{3}dx^{\mu}=\sin\theta_{+}d\varphi_{+}$
etc. for the original co-ordinates, and underlined flat indices for
the new complex co-ordinates: $x^{\underline{2}}=\mylam$, $x^{\underline{3}}=\myisom$.
Notice that the volume form is preserved by this change: $e^{2}\wedge e^{3}=e^{\underline{2}}\wedge e^{\underline{3}}$. 

The original real $S^{2}$ was the slice in which $\theta_{+},\varphi_{+}\in\mathbb{R}$,
but now we allow these to be complex. And importantly, we dualise
along the $\myisom$ direction, as well as time (and some of the torus
directions $y_{i}$). This gives dual metric 
\begin{equation}
ds_{AdS}^{\prime2}=-z^{2}dt^{2}+\frac{dz^{2}}{z^{2}},\qquad ds_{S^{+}}^{\prime2}=d\mylam^{2}+e^{-2i\mylam}d\myisom^{2}\label{eq:SS-new-metric}
\end{equation}
and a shift in the dilaton of 
\begin{equation}
\Phi''-\Phi'=\log\left(z\:e^{-i\mylam}\right)=\log\frac{z}{\cos\theta_{+}+i\sin\theta_{+}\sin\varphi_{+}}.\label{eq:SS-bose-deltaPhi}
\end{equation}
The dual metric $ds_{S^{+}}^{\prime2}$ in terms of $\theta_{+},\varphi_{+}$
is no longer real. To recover a sphere we must afterwards pick a different
real slice: define $\theta',\varphi'$ by \eqref{eq:SS-complex-coord}
with $\mylam$ replaced by $-\mylam$, i.e. 
\[
e^{-i\mylam}=\cos\theta'+i\sin\theta'\cos\varphi',\qquad\myisom=\frac{\sin\varphi'}{\cot\theta'+i\cos\varphi'}.
\]
Then obviously $ds_{S^{+}}^{\prime2}=d\theta^{\prime2}+\sin^{2}\theta'd\varphi^{\prime2}$.
And to recover $AdS_{2}$ we must also invert the radial co-ordinate,
defining $z'=1/z$, which clearly gives $ds_{AdS}^{\prime2}=(-dt^{2}+dz^{\prime2})/z^{\prime2}$. 

Returning to the fermionic T-duality, we now use very simple complex
combinations 
\[
\cE_{1}=E_{1}+iE_{4},\qquad\cE_{2}=-E_{2}+iE_{3}.
\]
For this to be meaningful we must specify the eigenvectors $\xi_{a}$
used; they are easy to calculate but messy to type. Let us order them
by the position of the first first nonzero component, and choose the
sign such that this is positive: 
\begin{equation}
(\xi_{a})^{\beta}=0\mbox{ for }\beta<a,\qquad(\xi_{a})^{\beta}>0\mbox{ for }\beta=a\label{eq:SS-ordering-the-xi}
\end{equation}
for example $\xi_{3}=\tfrac{1}{2\sqrt{2}}\big(0,0,\sqrt{1+\sqrt{\alpha}},0,0,\sqrt{1-\sqrt{\alpha}},0,0,\ldots\big)$.
Then we get 
\begin{equation}
\C=-2i\frac{\cos\theta_{+}+i\sin\theta_{+}\cos\varphi_{+}}{z}\left[\begin{array}{cc}
-e^{-i\varphi_{-}}\sin\theta_{-} & \cos\theta_{-}\\
\cos\theta_{-} & e^{i\varphi_{-}}\sin\theta_{-}
\end{array}\right]=\C_{+}\C_{-}\label{eq:SS-C}
\end{equation}
and hence the change in the dilaton is 
\begin{equation}
\Phi'-\Phi=\log\frac{\cos\theta_{+}+i\sin\theta_{+}\cos\varphi_{+}}{z}+\log(-2i)\label{eq:SS-fermi-deltaPhi}
\end{equation}
cancelling the bosonic shift perfectly (apart from a constant which
we could have absorbed). 

Next we must look at the change in the RR forms, and we begin with
the 2-form: 
\begin{align}
e^{i\Phi'}F^{(2)\prime} & =\sqrt{1-\alpha}\frac{(\sin\theta_{+}-i\cos\theta_{+}\cos\varphi_{+})e^{1}\wedge e^{2}+i\sin\varphi_{+}e^{1}\wedge e^{3}}{\cos\theta_{+}+i\sin\theta_{+}\cos\varphi_{+}}\nonumber \\
 & =-i\sqrt{1-\alpha}\,e^{1}\wedge e^{\underline{2}}\:.\label{eq:SS-F2prime}
\end{align}
On the second line we see that this simplifies greatly in terms of
the new complex co-ordinates (or rather, the corresponding flat directions).
The change in the 4-form field $\Delta F^{(4)}$ has real terms precisely
cancelling the original $F^{(4)}$, and imaginary terms which likewise
simplify greatly in the new co-ordinates, to give 
\begin{equation}
e^{i\Phi'}F^{(4)\prime}=-i\left(e^{1}e^{\underline{3}}e^{6}e^{8}+e^{1}e^{\underline{3}}e^{7}e^{9}\right)+i\sqrt{\alpha}\left(-e^{0}e^{\underline{2}}e^{6}e^{7}+e^{0}e^{\underline{2}}e^{8}e^{9}\right)+i\sqrt{1-\alpha}\left(e^{0}e^{\underline{3}}e^{4}e^{5}\right).\label{eq:SS-F4prime}
\end{equation}

Bosonic T-duality now acts on $F^{(2)\prime}$ and $F^{(4)\prime}$
in the usual way, and we dualise along directions $t=x^{0}$, $\myisom=x^{\underline{3}}$
and $x^{7}$, $x^{8}$. This gives $F^{(2)\prime\prime}=0$ and\newcommand{\etwo}{e^{\prime\underline{2}}}
\newcommand{\ethree}{e^{\prime\underline{3}}} 
\begin{align*}
F^{(4)\prime\prime} & =\left(-e^{0}e^{1}e^{6}e^{7}+e^{0}e^{1}e^{8}e^{9}\right)+\sqrt{\alpha}\left(\etwo\ethree e^{6}e^{8}+\etwo\ethree e^{7}e^{9}\right)+\sqrt{1-\alpha}\left(-e^{4}e^{5}e^{7}e^{8}+e^{4}e^{5}e^{6}e^{9}\right).
\end{align*}
We interpret these flat components as being with respect to the new
metric $ds^{\prime2}$ \eqref{eq:SS-new-metric} created simultaneously
by bosonic T-duality, hence $\ethree=e^{-i\mylam}d\myisom\,\neq\,e^{\underline{3}}$.
Note that $\etwo=-d\mylam=-e^{\underline{2}}$, so that the volume
form is preserved also by the second change of co-ordinates, to the
new real co-ordinates $\theta'=x^{2'},\varphi'=x^{3'}$: we write
$\etwo\wedge\ethree=e^{2'}\wedge e^{3'}$. Then we have recovered
the original flux \eqref{eq:AdS2-F4}.

\subsection{Aside on $AdS_{2}\times S^{2}\times T^{6}$ with other RR fluxes\label{sub:Aside-on-F2-F4-case}}

The $\alpha=1$ background in section \ref{sub:AdS2-simplest-case}
above has only $F^{(4)}$ turned on. Various other combinations of
fluxes are possible with the same geometry, and one of the cases considered
in \cite{Sorokin:2011rr} is 
\begin{equation}
\not F^{(2)}=-\Gamma^{01},\qquad\not F^{(4)}=\Gamma^{23}(\Gamma^{45}+\Gamma^{67}+\Gamma^{89}).\label{eq:AdS2-alt-F2-F4}
\end{equation}
This is also the case studied in \cite{Cagnazzo:2012uq}, and one
of the cases in \cite{Abbott:2015mla}. It is equally easy to show
the self-duality of this background using the same methods, and we
look at this briefly before moving on to $AdS_{3}\times M$. 

We can write 
\[
S=-4\Gamma^{01}\Gamma_{11}P,\qquad T=-4\Gamma^{01}\Gamma_{11}(1-P)
\]
in terms of 
\[
P=\tfrac{1}{4}(1-\Gamma^{6789}-\Gamma^{4589}-\Gamma^{4567}).
\]
The Killing spinors are still given by \eqref{eq:AdS2-simple-killing},
but the constraint on $\xi$ is now
\[
\Gamma^{0}\Gamma_{11}P\xi=\xi.
\]
We choose solutions $\xi_{i}$ which obey \eqref{eq:SS-ordering-the-xi},
and then take the following complex combinations:
\[
\cE_{1}=iE_{1}+E_{4},\qquad\cE_{2}=iE_{2}-E_{3}.
\]
This gives 
\[
\C=\frac{1}{z}\left[\begin{array}{cc}
ie^{i\phi_{+}}\sin\theta_{+} & -i\cos\theta_{+}\\
-i\cos\theta_{+} & -ie^{-i\phi_{+}}\sin\theta_{+}
\end{array}\right]
\]
and hence the change in the dilaton is $\Delta\Phi=-\log(z)$. This
is the same as \eqref{eq:AdS2-dilaton-fermishift} above, and is similarly
cancelled by the bosonic duality along time. 

The change in RR forms is
\begin{align*}
\Delta F^{(2)}= & \gamma^{01}\\
\Delta F^{(4)}= & -\gamma^{23}\left(\gamma^{45}+\gamma^{67}+\gamma^{89}\right)-i\,\gamma^{1468}+i\,\gamma^{1479}+i\,\gamma^{1569}+i\,\gamma^{1578}.
\end{align*}
Clearly the real terms again cancel the original flux. Acting with
bosonic T-duality on the remaining (imaginary) terms along directions
$t=x^{0}$, $x^{4}$, $x^{6}$ and $x^{8}$ returns us to \eqref{eq:AdS2-alt-F2-F4}.

\section{The background $AdS_{3}\times S^{3}\times S^{3}\times S^{1}$\label{sec:AdS3-Backgrounds}}

The metric here is still of the form \eqref{eq:defn-metric-1}, but
now with one more dimension. We choose nested co-ordinates for the
$S^{3}$ as in \cite{Lu:1998nu}: 
\begin{align*}
ds_{AdS_{3}}^{2} & =\frac{-dt^{2}+dx^{2}+dz^{2}}{z^{2}}\\
 & \qquad\Rightarrow\quad\omega_{t[02]}=\omega_{x[21]}=1/z\vphantom{1_{1_{1}}}\displaybreak[0]\\
ds_{S_{\pm}^{3}}^{2} & =d\theta_{\pm}^{2}+\sin^{2}\theta_{\pm}\left(d\varphi_{\pm}^{2}+\sin^{2}\varphi_{\pm}\:d\psi_{\pm}\right)\vphantom{1^{1^{1}}}\\
 & \qquad\Rightarrow\quad\omega_{\phi_{+}[34]}=-\cos\theta_{+},\quad\omega_{\psi_{+}[35]}=-\cos\theta_{+}\sin\varphi_{+},\quad\omega_{\psi_{+}[45]}=-\cos\varphi_{+}.
\end{align*}
We take the following flux \cite{Wulff:2014kja}: 
\begin{align}
\not F^{(4)} & =2\left(-\Gamma^{0129}+\sqrt{\alpha}\Gamma^{3459}+\sqrt{1-\alpha}\Gamma^{6789}\right)\label{eq:AdS3-F4}\\
 & =-4\Gamma^{0129}(1-P).\nonumber 
\end{align}
Bosonic T-duality along the $x^{9}$ direction returns us to the fluxes
of (for instance) \cite{Forini:2012bb}, and then taking the limit
$\alpha\to1$ reduces this to the IIB $AdS_{3}\times S^{3}\times T^{4}$
case for which \cite{OColgain:2012ca} studied fermionic T-duality.
Because of this study of the $\alpha=1$ case, we focus here only
on the $0<\alpha<1$ case. 

This background has 16 supersymmetries, and thus 8 Poincar\'e Killing
spinors, allowing for 4 complex directions along which we can perform
fermionic T-duality. We can build up a solution $E=z^{-1/2}R_{S+}R_{S-}\xi$
to the Killing spinor equation as before: 
\begin{itemize}
\item Poincar\'e killing spinors must now have $\partial_{t}E=0$ and $\partial_{x}E=0$.
The rest of the $\mu=t$ equation gives us a constraint $E=\Gamma^{019}PE$;
the $\mu=x$ equation is identical. The $\mu=z$ equation is solved
by $E\propto z^{-1/2}$.
\item When $\mu\in S_{+}$ we get (using $PE=E$) 
\[
D_{\mu}E=\frac{\sqrt{\alpha}}{2}\Gamma_{\mu}\Gamma^{01345}E
\]
which (again generalising \cite{Lu:1998nu}) is solved by a factor
$R_{S+}=e^{\frac{\theta_{+}}{2}\,\Gamma^{0145}}e^{\frac{\varphi_{+}}{2}\,\Gamma^{34}}e^{\frac{\psi_{+}}{2}\,\Gamma^{45}}.$
\item When $\mu\in S_{-}$ we get $D_{\mu}E=\frac{\sqrt{1-\alpha}}{2}\Gamma_{\mu}\Gamma^{01678}E.$
Notice that $[\Gamma_{\mu}\Gamma^{01678},R_{S+}]=[D_{\mu},R_{S+}]=0$,
so this is precisely the equation which must be solved by $R_{S-}$
alone. 
\end{itemize}
The complete solution is then\begin{small} 
\[
E(\xi)=\frac{1}{\sqrt{z}}\:\exp\Big(\frac{\theta_{+}}{2}\,\Gamma^{0145}\Big)\exp\Big(\frac{\varphi_{+}}{2}\,\Gamma^{34}\Big)\exp\Big(\frac{\psi_{+}}{2}\,\Gamma^{45}\Big)\:\exp\Big(\frac{\theta_{-}}{2}\,\Gamma^{0178}\Big)\exp\Big(\frac{\varphi_{-}}{2}\,\Gamma^{67}\Big)\exp\Big(\frac{\psi_{-}}{2}\,\Gamma^{78}\Big)\:\xi
\]
\end{small}with constant $\xi$ obeying 
\[
\Gamma^{019}\xi=P\xi=\xi.
\]
There are exactly 8 such spinors, as advertised. Let us order the
(orthogonal) $\xi_{a}$ such that 
\[
(\xi_{a})^{\beta}=0\mbox{ for }\beta<a,\qquad(\xi_{a})^{\beta}=\frac{1}{2}\mbox{ for }\beta=a.
\]
The orthogonality relation is now nontrivial for two values of $\mu$,
which in terms of $\myM^{\mu}$ defined in \eqref{eq:defn-V-orthog}
reads
\[
\myM^{t}=-1,\qquad\myM^{x}=-\sqrt{\alpha}\:\sigma^{2}\otimes\sigma^{2}\otimes1_{2\times2}-\sqrt{1-\alpha}\:1\otimes\sigma^{1}\otimes1\,.
\]
After some experiments we solve this by taking 
\[
\cE_{1}=E_{1}-iE_{8},\qquad\cE_{2}=E_{2}+iE_{7},\qquad\cE_{3}=E_{3}-iE_{6},\qquad\cE_{4}=E_{4}+iE_{5}.
\]
With this choice $\C$ factorises into 
\[
\C=\C_{+}(z,\theta_{+},\varphi_{+},\psi_{+})\:\C_{-}(\theta_{-},\varphi_{-},\psi_{-})
\]
where $\C_{+}=2i\,\sin\theta_{+}(\cos\varphi_{+}+i\sin\varphi_{+}\cos\psi_{+})/z$
is a number, and $\C_{-}$ a matrix with $\det\C_{-}=1$:\begin{small}
\[
\negthickspace\negthickspace\negthickspace\negthickspace\negthickspace\negthickspace\negthickspace\negthickspace\negthickspace\C_{-}=\left[\begin{array}{cccc}
-c_{\theta}-is_{\theta}s_{\varphi}(\hat{\alpha}c_{\psi}+\sqrt{\alpha}s_{\psi})\negthickspace\negthickspace & -i\hat{\alpha}c_{\varphi}s_{\theta} & -\sqrt{\alpha}c_{\varphi}s_{\theta} & s_{\theta}s_{\varphi}(\sqrt{\alpha}c_{\psi}-\hat{\alpha}s_{\psi})\\
-i\hat{\alpha}c_{\varphi}s_{\theta} & \negthickspace\negthickspace is_{\theta}s_{\varphi}(\hat{\alpha}c_{\psi}-\sqrt{\alpha}s_{\psi})-c_{\theta}\negthickspace & s_{\theta}s_{\varphi}(\sqrt{\alpha}c_{\psi}+\hat{\alpha}s_{\psi}) & \sqrt{\alpha}c_{\varphi}s_{\theta}\\
-\sqrt{\alpha}c_{\varphi}s_{\theta} & s_{\theta}s_{\varphi}(\sqrt{\alpha}c_{\psi}+\hat{\alpha}s_{\psi}) & \negthickspace c_{\theta}+is_{\theta}s_{\varphi}(\hat{\alpha}c_{\psi}-\sqrt{\alpha}s_{\psi})\negthickspace\negthickspace & i\hat{\alpha}c_{\varphi}s_{\theta}\\
s_{\theta}s_{\varphi}(\sqrt{\alpha}c_{\psi}-\hat{\alpha}s_{\psi}) & \sqrt{\alpha}c_{\varphi}s_{\theta} & i\hat{\alpha}c_{\varphi}s_{\theta} & \negthickspace\negthickspace c_{\theta}-is_{\theta}s_{\varphi}(\hat{\alpha}c_{\psi}+\sqrt{\alpha}s_{\psi})
\end{array}\right]
\]
\end{small}writing (only here) $s_{\theta}=\sin\theta_{-}$, $c_{\theta}=\cos\theta_{-}$
and $\hat{\alpha}=\sqrt{1-\alpha}$. Then we get 
\begin{equation}
\Phi'-\Phi=2\log\frac{2\sin\theta_{+}(\cos\varphi_{+}+i\sin\varphi_{+}\cos\psi_{+})}{z}.\label{eq:AdS3-fermishift}
\end{equation}

To undo this shift using bosonic T-duality we again need to dualise
along some complex directions. The parameterisation of the coset element
chosen in \cite{Abbott:2015mla}  implies a metric for $S_{+}^{3}$
generalising \eqref{eq:metric-S2-lambda}: 
\[
ds_{S_{+}^{3}}^{2}=d\lambda_{3}^{2}+e^{2i\lambda_{3}}d\lambda_{+}^{2}-e^{2i\lambda_{3}}d\lambda_{-}^{2}.
\]
Dualising along $\lambda_{+}$ and $\lambda_{-}$ (as well as two
$AdS_{3}$ directions $t,x$) gives a shift in the dilaton of 
\[
\Phi''-\Phi'=2\log\left(z\,e^{-i\lambda_{3}}\right)
\]
from which we read off 
\[
e^{i\lambda_{3}}=\sin\theta_{+}(\cos\varphi_{+}+i\sin\varphi_{+}\cos\psi_{+}).
\]
The same duality also changes the metric $ds^{2}\to d\lambda_{3}^{2}+e^{-2i\lambda_{3}}d\lambda_{+}^{2}-e^{+2i\lambda_{3}}d\lambda_{-}^{2}$,
which as before we can absorb into the relation between $\lambda_{3}$
and the real co-ordinates for $S^{3}$, writing $\smash{e^{-i\lambda_{3}}}=\sin\theta'(\cos\varphi'+i\sin\varphi'\cos\psi')$. 

Finally we look at the change in the RR fields. Fermionic T-duality
produces a 2-form field which, as in \eqref{eq:SS-F2prime} above,
is much simpler written in the new complex co-ordinates. Writing $e^{\underline{3}}=d\lambda_{3}$,
this is 
\begin{align}
\Delta F^{(4)} & =2\sqrt{1-\alpha}\left\{ \cot\theta_{+}e^{2}e^{3}+\frac{i(\cos\varphi_{+}\cos\psi_{+}+i\sin\varphi_{+})\,e^{2}e^{4}-i\sin\psi_{+}e^{2}e^{5}}{\sin\theta_{+}(\cos\varphi_{+}+i\sin\varphi_{+}\cos\psi_{+})}\right\} \nonumber \\
 & =2i\sqrt{1-\alpha}\:e^{2}e^{\underline{3}}\,.\label{eq:AdS3-delta-F2}
\end{align}
One of the $\sqrt{\alpha}$ terms in the change in the 4-form field
is similarly simple, on the first line here: 
\begin{align*}
\Delta F^{(4)} & =-2\sqrt{\alpha}\:e^{0}e^{1}e^{9}\left[\cot\theta_{+}e^{3}+\frac{i(\cos\varphi_{+}\cos\psi_{+}+i\sin\varphi_{+})\,e^{4}-i\sin\psi_{+}e^{5}}{\sin\theta_{+}(\cos\varphi_{+}+i\sin\varphi_{+}\cos\psi_{+})}\right]\\
 & \qquad+2\sqrt{\alpha}\:e^{3}e^{4}e^{5}e^{9}+2\sqrt{1-\alpha}\:e^{6}e^{7}e^{8}e^{9}-2\,e^{0}e^{1}e^{2}e^{9}\\
 & \qquad+2\,e^{2}e^{9}\left[\cot\theta_{+}e^{4}e^{5}+\frac{-i(\cos\varphi_{+}\cos\psi_{+}+i\sin\varphi_{+})\,e^{3}e^{5}-i\sin\psi_{+}e^{3}e^{4}}{\sin\theta_{+}(\cos\varphi_{+}+i\sin\varphi_{+}\cos\psi_{+})}\right].
\end{align*}
On the second line we have terms cancelling the original $F^{(4)}$.
The square bracket on the third line here is precisely $i\,e^{\underline{4}}e^{\underline{5}}$,
as can be seen by noting that the volume form should be the same as
for the initial co-ordinates --- the product of the two square brackets
above is 
\[
\left[i\,e^{\underline{3}}\right]\wedge\left[i\,e^{\underline{4}}\wedge e^{\underline{5}}\right]=\left(\cot^{2}\theta_{+}+\ldots\right)e^{3}\wedge e^{4}\wedge e^{5}=-e^{3}\wedge e^{4}\wedge e^{5}.
\]
Thus the RR fields after the fermionic step are: 
\begin{equation}
\begin{aligned}e^{\Phi'}F^{(2)\prime} & =2i\sqrt{1-\alpha}\:e^{2}e^{\underline{3}}\\
e^{\Phi'}F^{(4)\prime} & =2i\sqrt{\alpha}\:e^{0}e^{1}e^{\underline{3}}e^{9}+2i\,e^{2}e^{\underline{4}}e^{\underline{5}}e^{9}.
\end{aligned}
\label{eq:AdS3-all-Fprime}
\end{equation}
Bosonic T-duality along $t,\,x,$ and $\lambda_{\pm}$ (i.e. directions
0,1,$\underline{4}$,$\underline{5}$) leads to 
\[
e^{\Phi''}F^{(4)\prime\prime}=-2e^{0}e^{1}e^{2}e^{9}+2\sqrt{\alpha}\:e^{\prime\underline{3}}e^{\prime\underline{4}}e^{\prime\underline{5}}e^{9}+2\sqrt{1-\alpha}\:e^{6}e^{7}e^{8}e^{9}.
\]
Then using $e^{\prime\underline{3}}e^{\prime\underline{4}}e^{\prime\underline{5}}=e^{3'}e^{4'}e^{5'}$
to write this in terms of the final set of real co-ordinates $\theta',\varphi',\psi'$,
we recover the original RR field, \eqref{eq:AdS3-F4}.

\section{Conclusions\label{sec:Conclusions}}

In this paper we have shown that type IIA string theory on backgrounds
$AdS_{2}\times S^{2}\times T^{6}$ and $AdS_{d}\times S_{+}^{d}\times S_{-}^{d}\times T^{10-3d}$
for $d=2,3$ is self-dual under bosonic and fermionic T-duality. We
have deliberately kept our computations as `low-tech' as possible,
constructing Killing spinors by hand following \cite{Lu:1998nu},
and then using the Buscher rules of \cite{Berkovits:2008ic} to work
out the effect of fermionic T-duality. This approach makes particularly
clear that we cannot avoid bosonic T-duality along torus directions,
and in the $S_{+}\times S_{-}$ cases also along some complexified
sphere directions. 

All of these backgrounds have also been studied (in various ways)
using coset sigma models. These approaches have different strengths
and weaknesses: 
\begin{itemize}
\item $AdS_{3}\times S^{3}$ is the bosonic part of $PSU(1,1\vert2)^{2}/SU(1,1)\times SU(2)$,
which in turn is a truncation of the $PSU(2,2\vert4)$ super-coset
used for $AdS_{5}\times S^{5}$. This truncation lets you re-use many
parts of the story, however it omits the flat torus directions, and
half the fermions of the GS action.\footnote{In the integrable picture, the same truncation of $AdS_{5}\times S^{5}$
omits the massless modes, from the torus directions and their superpartners.
This was the first sector to be explored, see review \cite{Sfondrini:2014via}.
Many details of the massless sector are still being worked out \cite{Borsato:2014hja,Sundin:2015uva}. } It is possible to fix a kappa-gauge such that the non-coset fermions
all vanish, as was done by \cite{Adam:2009kt}. The approach of \cite{Abbott:2015mla}
was to start with this coset, and then explicitly add the non-coset
fermions. This made visible the need for T-duality along torus directions. 
\item $AdS_{3}\times S^{3}\times S^{3}$ is likewise the bosonic part of
the exceptional group coset $D(2,1;\alpha)^{2}/SO(1,2)\times SO(3)^{2}$,
which again omits half the GS fermions.\footnote{However in this case the coset only omits two bosonic directions,
see \cite{Babichenko:2009dk,Lloyd:2013wza,Abbott:2014rca}.} For this case the approach of \cite{Abbott:2015mla} was different:
since all the Killing spinors lie in the coset directions, one can
re-cast the fermionic Buscher rules into this language. Once this
is done, self-duality follows in a few lines. 
\item In the case $AdS_{2}\times S^{2}\subset PSU(1,1\vert2)/SO(1,1)\times U(1)$,
only 8 of the GS string's 32 fermions are described by these coset,
thus a kappa gauge cannot remove all the non-coset fermions. (This
is also true for $AdS_{2}\times S^{2}\times S^{2}\subset D(2,1;\alpha)/SO(1,1)\times SO(2)^{2}$.)
The same list of approaches is possible here: one can truncate to
the coset \cite{Adam:2009kt}, or explicitly add non-coset fermions
to the action \cite{Abbott:2015mla}, or generate Killing spinors
from the coset to use in the supergravity rules \cite{Abbott:2015mla}. 
\end{itemize}
Apart from these backgrounds, in the introduction we mentioned the
goal of resolving the longstanding problem with $AdS_{4}\times CP^{3}$
self-duality \cite{Adam:2009kt,Adam:2010hh,Bakhmatov:2010fp}. Another
open direction concerns backgrounds with mixed RR and NS-NS fields.
All of the above geometries can be supported by some such mixture,
and some can be supported by pure NS-NS flux \cite{Wulff:2014kja}.
The integrability of such backgrounds has been a topic of much recent
interest \cite{Babichenko:2014yaa,Lloyd:2014bsa} but nothing is known
about their fermionic T-duality.

\subsection*{Acknowledgements}

We have benefited greatly from discussions with our co-authors of
\cite{Abbott:2015mla}, and more locally with Emanuel Malek. We must
especially thank Martin Wolf, for pointing out to us the need to dualise
along complex $\lambda_{\pm}$. 

Michael is supported by an NRF Innovation Fellowship. Jeff is supported
by the NRF CPRR program under GUN 87667. Justine thanks the UCT Faculty
of Science for Ph.D. support. 

\appendix

\section{Bosonic T-duality\label{sec:Bosonic-T-duality}}

While our focus here is on fermionic T-duality, the self-duality involves
also bosonic T-duality. For this the tools are much better developed,
and we review briefly what we need of these.

\subsection{Generalised metric and dilaton}

The most convenient formalism for working out the effect of bosonic
T-duality on the metric $G$, the Kalb--Ramond antisymmetric $B$
field, and the dilaton $\Phi$ is to use generalised geometry, where
T-duality is generated by an $O(D,D)$ transformation. 

The generalised metric $M$ and generalised dilaton $d$ are defined
by 
\[
M=\left[\begin{array}{cc}
G-BG^{-1}B & -BG^{-1}\\
G^{-1}B & G^{-1}
\end{array}\right],\qquad e^{-2d}=e^{-2\Phi}\sqrt{\det G}.
\]
$O(D,D)$ transformations act $M\to M'=T^{-1}M\,T$, and $d$ is invariant.
T-duality is implemented by the following matrix, with for instance
$v=(0,\ldots,0,1)$ to dualise along the last direction: 
\[
T_{v}=\left[\begin{array}{cc}
1-n & n\\
n & 1-n
\end{array}\right],\qquad n_{\mu\nu}=v_{\mu}v_{\nu}.
\]
Then it is easy to see that dualising along several co-ordinate directions,
we have $T_{v_{1}}T_{v_{2}}=T_{v_{1}+v_{2}}$. Reading off $G^{\prime-1}$
from $M'$, the change in the dilaton is clearly 
\begin{equation}
\Delta\Phi=\frac{1}{4}\log\frac{\det G'}{\det G}.\label{eq:rules-bose-deltaPhi}
\end{equation}
When $B=0$, if we write $g_{\mu\nu}$ for the block of $G_{\mu\nu}$
in the directions along which we are dualising, then this becomes
simply 
\[
\Delta\Phi=-\frac{1}{2}\log(\det g).
\]

\subsection{Effect on Ramond--Ramond fields}

One way to work out the changes in the RR form fields, from Fukuma
et al. \cite{Fukuma:1999jt}, begins by writing them using a set of
fermionic creation operators $\psi^{m}$. In terms of the potentials
$C^{(p)}$ this reads 
\[
\ket{C}=\sum_{p}\frac{1}{p!}C_{mn\cdots p}^{(p)}\psi^{m}\psi^{n}\cdots\psi^{p}\ket{0}
\]
where the algebra is 
\[
\{\psi_{m},\psi^{n}\}=\delta_{m}^{n},\qquad\{\psi^{m},\psi^{n}\}=\{\psi_{m},\psi_{n}\}=0
\]
and destruction operators $\psi_{m}$ annihilate the vacuum $\ket{0}$.
Then the action of bosonic T-duality in the $i$ direction is given
by 
\begin{equation}
T_{m}=\psi^{m}+\psi_{m}\,.\label{eq:rules-bose-Fukuma}
\end{equation}
Note that different $T_{m}$ clearly anti-commute, so the sign of
the final $C^{(p)\prime}$ depends on the order (but is not physical).
However there are two additions to this which matter for us: 
\begin{itemize}
\item If there is a nontrivial dilaton, these rules should apply instead
to $e^{\Phi}dC^{(p)}$. This is mentioned in the very last paragraph
of \cite{Fukuma:1999jt} (before the appendix), but is also implicit
in \cite{Berkovits:2008ic,OColgain:2012ca}. 
\item T-duality along the time direction exchanges real and imaginary fluxes
\cite{Hull:1998vg}, which means that \eqref{eq:rules-bose-Fukuma}
applies to $m\neq0$, and $T_{0}=i(\psi^{0}+\psi_{0})$. 
\item Also note that if there is a $B$-field turned on, then these rules
should apply not to $C$ but to the modified potential $D$ introduced
by \cite{Fukuma:1999jt}. We do not consider this case. 
\end{itemize}
Another way to write this was given by Hassan \cite{Hassan:1999bv},
acting directly on the fields $F^{(p)}$. When dualising along direction
$x^{9}$, the new fluxes are 
\begin{equation}
\begin{aligned}F_{9no\cdots q}^{(p)\prime} & =F_{no\cdots q}^{(p-1)}-\frac{p-1}{G_{99}}G_{9[n}^{\phantom{()}}F_{9o\cdots q]}^{(p-1)}\\
F_{mn\cdots q}^{(p)\prime} & =F_{9mn\cdots q}^{(p+1)}-pB_{9[n}^{\phantom{()}}F_{o\cdots q]}^{(p)\prime}
\end{aligned}
\label{eq:rules-bose-Hassan}
\end{equation}
with $m,n,\ldots q=0,1,\ldots8$, and antisymmetrisation $[n\quad9,o,\ldots q]$
acting on the letters only. In all our cases $G$ is diagonal and
$B$ is zero, and thus the second terms here vanish. Note that when
dualising along time we still need to insert a factor of $i$ \cite{Hull:1998vg}.
And with a nontrivial dilaton we should include its factor too. 

We can write the transformation very simply in terms of the bispinor
defined in \eqref{eq:IIA-bispinor-defn} above, and its IIB cousin
\begin{equation}
F^{\alpha\beta}=F_{m}^{(1)}(\gamma^{m})^{\alpha\beta}+\frac{1}{3!}\:F_{mnp}^{(3)}\:(\gamma^{m})^{\alpha\gamma}(\gamma^{n})_{\gamma\delta}(\gamma^{p})^{\delta\beta}+\frac{1}{2\times4!}\:F_{mnopq}^{(5)}\:(\gamma^{m}\gamma^{n}\gamma^{o}\gamma^{p}\gamma^{q})^{\alpha\beta}.\label{eq:IIB-bispinor-defn}
\end{equation}
Incorporating $e^{\Phi}$ as above, we then have\footnote{Notice that \eqref{eq:rules-bose-Fukuma} and \eqref{eq:rules-bose-Hassan}
appear to produce fields $F^{(p)}$ with $p>5$. These are of course
not independent, and the IIA ones are related by \cite{Fukuma:1999jt}
\[
*F^{(8)}=F^{(2)},\qquad*F^{(6)}=-F^{(4)}.
\]
This fact is built into the bispinor \eqref{eq:IIA-bispinor-defn},
where it amounts to $\prod_{m=0}^{9}\gamma^{m}=1$. } 
\begin{equation}
\begin{aligned}T^{m}:\quad e^{\Phi'}F_{\alpha\beta}^{\prime} & =e^{\Phi}(\gamma^{m})_{\alpha\delta}F_{\;\;\beta}^{\delta},\qquad\;\,\mbox{or}\qquad e^{\Phi'}F_{\;\;\,\beta}^{\prime\alpha}=e^{\Phi}(\gamma^{m})^{\alpha\delta}F_{\delta\beta},\qquad m\neq0\\
T^{0}:\quad e^{\Phi'}F_{\alpha\beta}^{\prime} & =i\,e^{\Phi}(-\delta_{\alpha\delta})F_{\;\;\beta}^{\delta},\qquad\mbox{or}\qquad e^{\Phi'}F_{\;\;\,\beta}^{\prime\alpha}=i\,e^{\Phi}(\delta^{\alpha\delta})F_{\delta\beta}.
\end{aligned}
\label{eq:rules-F}
\end{equation}
In fact we always need four dualities, and acting on the $F'$ produced
by fermionic T-duality this reads 
\[
e^{\Phi''}F''=\pm i\gamma^{0mnp}(e^{\Phi'}F')
\]
where $m,n,p$ vary in the different backgrounds. (We have always
chosen the sign $\pm$, or equivalently the order of the dualities,
to produce the tidiest result.)

\section{Gamma Matrices\label{sec:Gamma-Matrices}}

We adopt the same conventions as \cite{Bakhmatov:2011ab,OColgain:2012ca}
among others, defining 
\[
\Gamma^{0}=\left[\begin{array}{cc}
0 & 1\\
-1 & 0
\end{array}\right],\qquad\Gamma^{m}=\left[\begin{array}{cc}
0 & (\gamma^{m})^{\alpha\beta}\\
(\gamma^{m})_{\alpha\beta} & 0
\end{array}\right]
\]
with $16\times16$ blocks as follows: \newcommand{\foursigma}[4]{\sigma^{#1} \otimes \sigma^{#2} \otimes \sigma^{#3} \otimes \sigma^{#4}}
\newcommand{\threea}[4]{\sigma^{#1} \otimes \myone \otimes \sigma^{#3} \otimes \sigma^{#4}}
\newcommand{\threeb}[4]{\sigma^{#1} \otimes \sigma^{#2} \otimes \myone \otimes \sigma^{#4}}
\newcommand{\threec}[4]{\sigma^{#1} \otimes \sigma^{#2} \otimes \sigma^{#3} \otimes \myone}
\newcommand{\onesigma}[1]{\sigma^{#1} \otimes \myone \otimes \myone \otimes \myone}
\newcommand{\myone}{\, 1 \, }
\begin{align*}
\gamma^{1} & =\foursigma{2}{2}{2}{2} & \gamma^{6} & =\threeb{2}{2}{0}{1}\\
\gamma^{2} & =\threea{2}{0}{1}{2} & \gamma^{7} & =\threeb{2}{2}{0}{3}\displaybreak[0]\\
\gamma^{3} & =\threea{2}{0}{3}{2} & \gamma^{8} & =\onesigma{2}\\
\gamma^{4} & =\threec{2}{1}{2}{0} & \gamma^{9} & =\onesigma{3}\\
\gamma^{5} & =\threec{2}{3}{2}{0}
\end{align*}
and $(\gamma^{0})_{\alpha\beta}=-\delta_{\alpha\beta}$, $(\gamma^{0})^{\alpha\beta}=\delta^{\alpha\beta}$.
These $\gamma^{m}$ are all clearly real and symmetric. We take the
charge conjugation matrix to be $C=\Gamma^{0}$. And we notice that
$\Gamma_{11}\equiv\prod_{m=0}^{9}\Gamma^{m}=\sigma^{3}\otimes1_{16\times16}$.

\bibliographystyle{my-JHEP-5}
\bibliography{/Users/me/Documents/Papers/complete-library-processed,complete-library-processed}

\end{document}